\documentclass[epjST]{svjour}
\usepackage{graphics}
\begin{document}

\title{The clash of the Titans: how preferential attachment helps the survival of the smallest}

\author{Suman Aich\thanks{\email{suman12t@bose.res.in}}  \and Anita Mehta\thanks{\email{anita@bose.res.in}}}

\institute{Department of Theoretical Sciences, Satyendra Nath Bose National Centre for Basic Sciences, Block JD, Sector-III, Salt Lake, Kolkata-700098, India}

\abstract{
We examine the effects of preferential attachment on a model of competing clusters. In the original model, cluster masses grow at the expense of their neighbours; on a lattice, this is known to result in the asymptotic survival and indefinite growth of clusters which are isolated from each other. The presence of preferential attachment results in an inhomogeneous topology, where hubs monopolise the connections, while most other nodes are sparsely connected. Interestingly, this results in the protection of the less massive clusters from annihilation, to which the hubs are doomed.
}

%%%%%%%%%%%%%%%%--------------------------------------------------------------------

\maketitle

%%%%%%%%%%%%%%%%%%%%%%%%%%%%%%%%%%%%%%%%%%%%%%%%%%%%%%%%%%%%%%%%%%%%%%%%%%%%%%%%%%%%%%%%%%%%%%%%%%%%%%%%
\section{Introduction}
\label{intro}

Many models in non-equilibrium physics are concerned with predator-prey like interactions where, typically, larger entities feed off smaller ones \cite{KT}. Such competitive dynamics model the survival of the fittest, known to be a strong candidate for evolutionary growth \cite{SF1,SF2}. In many cases, though, the competition occurs in the presence of a medium, which has its own dynamics: Ref. \cite{TM} provides an example, where an active nutrient pool affects the dynamics of an interacting microbial population. Another practical example concerns traders who interact in the presence of a central reserve \cite{RGR}. The latter is actually based on the model of Ref. \cite{JML}, in which the dynamics of competing clusters in the presence of a medium were extensively studied; this model also forms the basis of our present work.

In earlier work \cite{NNT}, the model of \cite{JML} was studied on random networks. To this we have added preferential attachment, which is known to govern realistic networks such as the World Wide Web \cite{r5,BAREV,r6}, movie actor collaboration networks\cite{r7,r8,r9}, science collaboration networks \cite{r10,r11,r12}, cellular networks \cite{r13}, and ecological networks \cite{r14}. 

The connections within a real system are in general not random; these are often deliberate, in which an agent may want to be connected with those who are the most popular, with the highest connectivity. It is for this reason that we incorporate preferential attachment in this paper (PA(I)) \cite{BAREV,BBASI1}. There is often also the tendency for highly connected agents to connect with those of equally high connectivity (PA(II)) \cite{BAREV,BBASI2}; this scenario is also investigated in this paper. The use of PA(I) is expected to give a topology similar to that of Model B, which is a limiting case of the Barab\'{a}si-Albert model \cite{BAREV}.

The plan of this paper is as follows. Starting with an outline of the model \cite{JML}, the different processes of extra wiring which we are going to use are described. The following section discusses the degree distributions and survival probabilities at different connection probabilities $p$ $(0\leq p \leq 1)$ under different wiring processes. Next, the dependence of the survival of a cluster on its degree is investigated. Additionally, the reduced time $\tau$ for isolation of eventual survivors and the number of eventual survivors $N_{ev}$ are plotted against $p$ for different wiring scenarios. Finally, the cumulative mass difference distributions for all the clusters and for the eventual survivors are found, as well as the fraction of the clusters $\alpha_{ev}$ which survive against the odds. We discuss our results in the last section.

%%%%%%%%%%%%%%%%%%%%%%%%%%%%%%%%%%%%%%%%%%%%%%%%%%%%%%%%%%%%%%%%%%%%%%%%%%%%%%%%%%%%%%%%%%%%%%%%%%%%%%%%%%%%%%%%%%%%
\section{ A review of the competitive cluster model }
\label{sec:1}

Throughout this paper, clusters are situated at the sites of a regular square lattice, connected to four nearest neighbours apiece; these clusters are fully characterized by their masses. The evolution of their time-dependent masses m(t) is governed by \cite{JML,NNT}:

\begin{equation}
\frac{dm_i}{dt}={\Big ( }\frac{\alpha}{t}-\frac{1}{t^{1/2}} \sum_{i \neq j} {g_{ij}} \frac{dm_j}{dt} {\Big )}m_i - \frac{1}{m_i} ,
\end{equation}
for the $i$-th cluster, where, $\alpha(>1/2)$ is called the accretion parameter; $ g_{ij} $, being very small, denotes the interaction strength between the $i$-th and $j$-th clusters. Here, the first term in the R.H.S. of Eq. (1) represents the gain term of the $i$-th cluster; the second term indicates the interaction of $i$-th cluster with its neighbours and the last term (loss term) is due to the dissipation of the $i$-th cluster to the surroundings. Throughout this paper, $\alpha$ and $g$ are set to be equal to $1.0$ and $10^{-3} $, respectively.

\begin{figure}
\centering
\resizebox{0.75\columnwidth}{!}{
\includegraphics{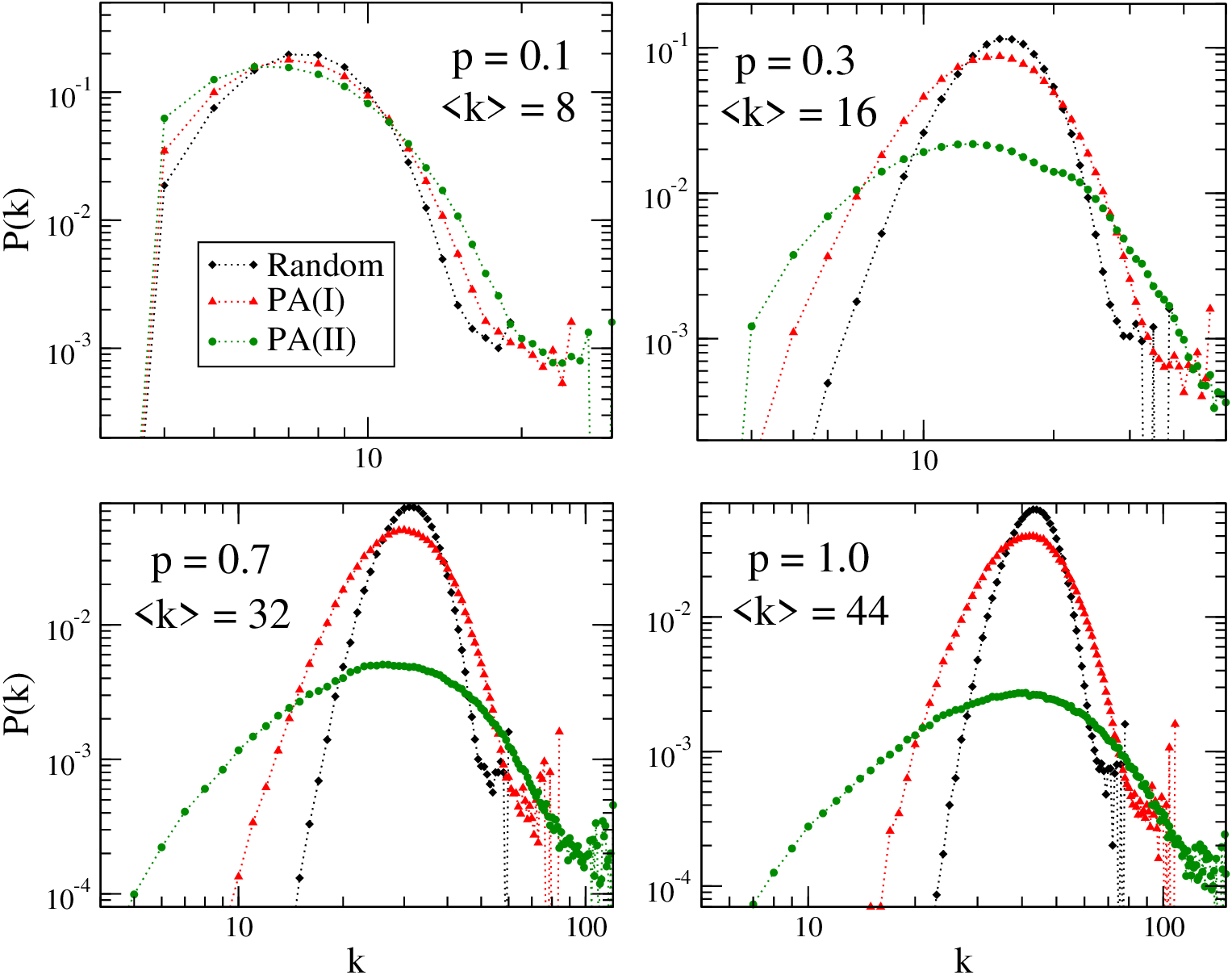} }
\caption{(Color online). Plot of the degree distribution for a $25 \times 25$ lattice system under a 20-cycle scheme for the different values of $p$ at three wiring scenarios. At a lower $p$ value these three plots almost merge. But, for higher $p$ the plots are separated out with PA(II) producing the flattest and the random case the sharpest peak, with  PA(I) in between. At high $p$, PA(II) has the most pronounced hubs. At $ p = 1.0 $, all of the plots have their peaks about $\langle k \rangle$. }
\end{figure}

Now, writing the time $t$ and mass $m$ as,

\begin{center}
$ s=ln(\frac{t}{t_0}) $ (reduced time),
~\\
%\newline
$ x_i=\frac{m_i}{t^{1/2}} $ (reduced mass),
\end{center}
respectively, Eq. (1) reduces to:
\begin{equation}
x_i^{'} ={\Big (}\frac{2\alpha-1}{2}-\sum_{i \neq j} {g_{ij} {\big(}\frac{x_j}{2}+x_j^{'}{\big)}}{\Big )}x_i-\frac{1}{x_i} .
\end{equation}
Next, the time variation of the reduced mass for the $n$-th cluster can be written as \cite{JML}:
\begin{equation}
x_n^{'} ={\Big (}\frac{2\alpha-1}{2}+g\sum_{k} {{\big (}\frac{1}{x_k}-\alpha x_k{\big )}}{\Big )}x_n-\frac{1}{x_n} ,
\end{equation}
where $k$ runs over the all nearest neighbours of the $n$-th node \cite{JML}.

Some clusters die, as a result of this dynamics; other, special clusters become eventual survivors with their masses growing forever. The survival probability ($S(s)$) for the clusters corresponds to the ratio of the number of surviving clusters at reduced time $s$ to the total number of initial clusters. If $S(s)$ is plotted as a function of $s$, it shows two distinct stages \cite{JML}. Stage I, which shows fast dynamics, is characterized by independent interactions; clusters having initial masses below some threshold $x_*$ (i.e., $x_i<x_*$), die out without any interaction among themselves and $S$ reaches a plateau value $S_{(1)}$. The critical mass $x_*$ is given by $x_* = \surd{ \frac{2}{2\alpha-1}} $ and here, $x_* = \surd{2}$ for $\alpha = 1.0$. Stage II consists of a slow dynamics  where clusters which have survived Stage I interact collectively with their nearest neighbours. When the dynamics are complete, with eventual survivors being isolated, $S$ attains $ S_{(\infty)} $ \cite{JML}.

%%%%%%%%%%%%%%%%%%%%%%%%%%%%%%%%%%%%%%%%%%%%%%%%%%%%%%%%%%%%%%%%%%%%%%%%%%%%%%%%%%%%%%%%%%%%%%%%%%%%%%%%

\section{Additional wiring strategies}
\label{sec:2}

\begin{figure}
\centering
\resizebox{0.7\columnwidth}{!}{
\includegraphics{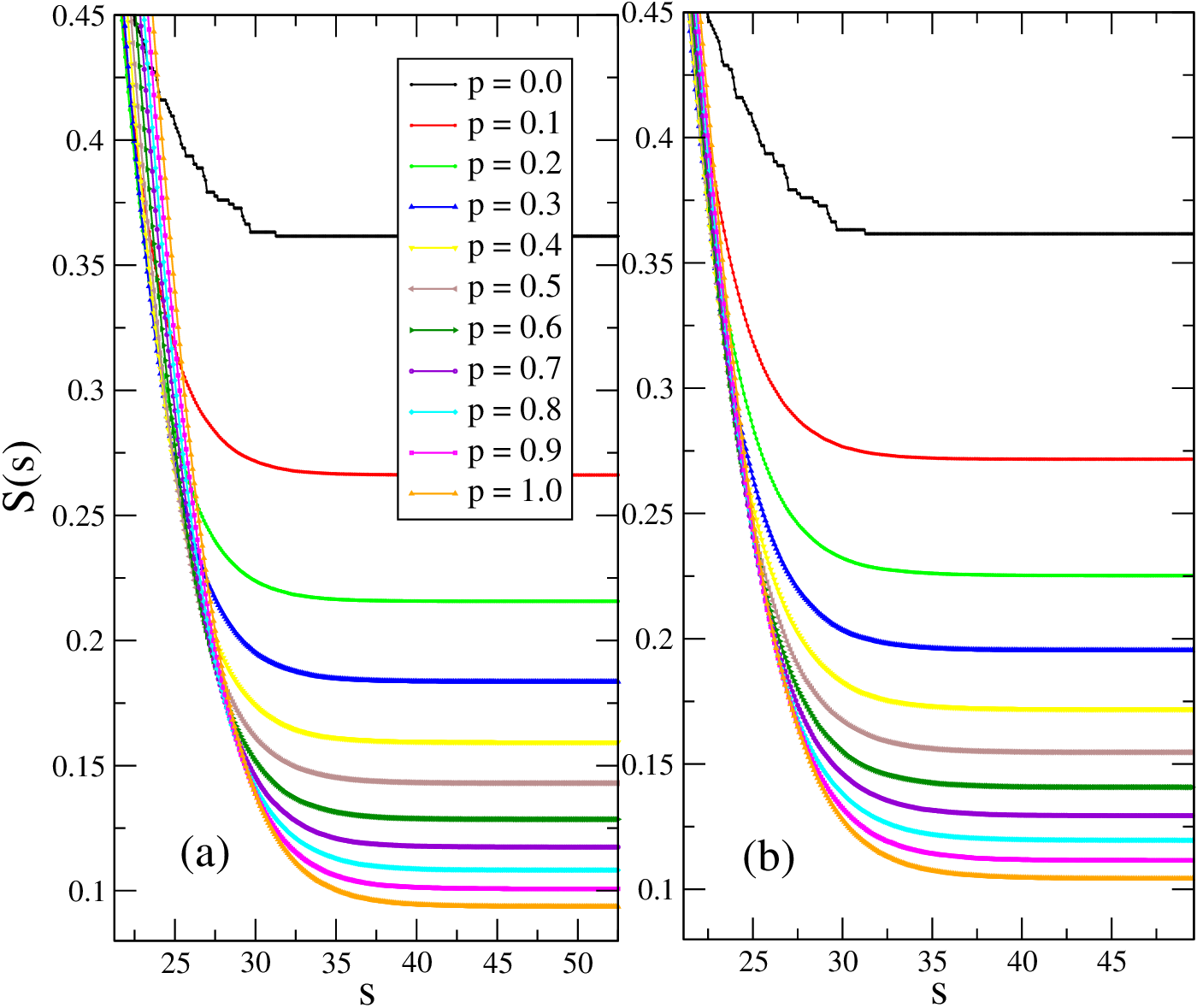} }
\caption{(Color online). Variation of the survival probability $S(s)$ with the reduced time $s$ in Stage II for two scenarios of extra wiring in a $25 \times 25$ lattice under the 20-cycle scheme ($S_{(1)} = 0.79$). In each cycle, for a particular value of $p$ the total $25 \times 25 \times p$ numbers of extra wiring are added in the whole lattice without any overlapping of edges and self-connections. In the 20-cycle scheme, the same thing is done twenty times over the whole lattice. (a) Curves resulting from the random wiring show two stages. (b) Curves obtained for PA(I) show the same two stages.}
\end{figure}

In a square lattice, each node is connected with its four nearest neighbours. In addition to that, extra connections are introduced among these nodes without any overlap of edges and avoiding self-connections. Extra edges can be introduced in the lattice by following the following strategies:

{\bf{(i) Random Wiring: }} In this method, both end points of an extra edge are selected randomly \cite{ER1,ER2,ER3}. As a result, most nodes will show connectivities closely distributed about the average, with few large fluctuations.

{\bf{(ii) Wiring by Preferential Attachment(I): }} Here, the first node is chosen randomly, and the other end of the edge is selected with preferential attachment: 
$$\Pi(k_i) = \frac{k_i}{ \sum_j k_j},$$ where $k_i$ is the degree of $i$-th node and $\Pi (k_i)$ is its probability to be connected with the randomly chosen node \cite{BAREV,BBASI1}. In this case, nodes with rather high, as well as low, connectivities are formed.

{\bf{(iii) Wiring by Preferential Attachment(II): }} Here, pairs of nodes with high degrees are preferentially chosen for connection. This nature of attachment probability is shown by the co-authorship network (Barabasi et al., 2001) \cite{BAREV,BBASI2}. Here, the connection probability $\Pi_{ij}$ of $i$-th and $j$-th node is such that, $$\Pi_{ij} \sim k_ik_j,$$ where $k_i$ is the degree of the $i$-th node.

In the numerical simulations, the additional wiring probability $p$ $(0\leq p \leq 1)$ corresponds to $Np$ extra wirings, where $N = L^2$ for a $L \times L$ system. Moreover, an $m$-cycle scheme corresponds to $Np$ extra wirings, added $m$ times throughout the whole lattice without any overlap and self-connections.

\begin{figure}
\centering
\resizebox{0.6\columnwidth}{!}{
\includegraphics{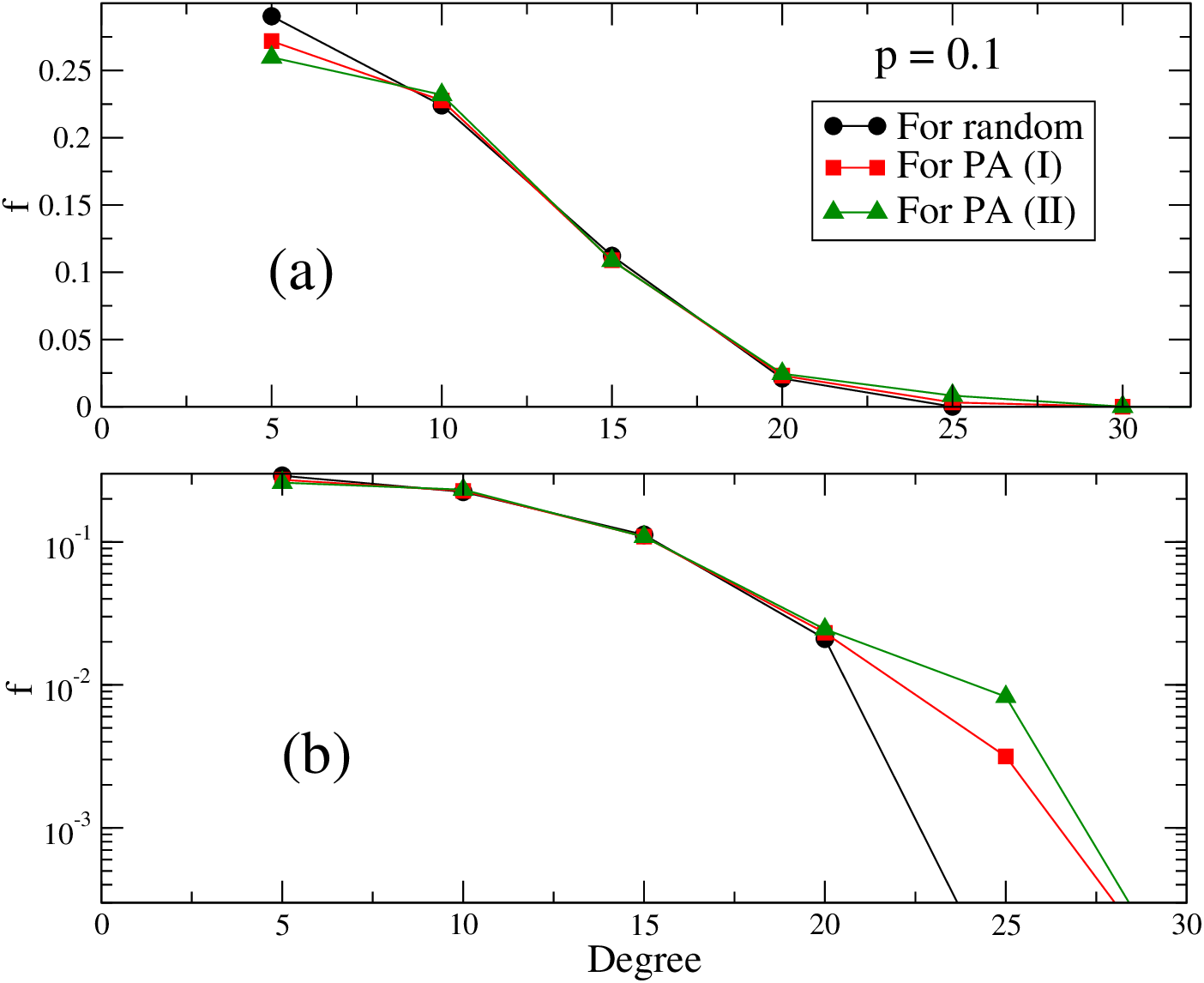} }
\caption{(Color online). (a) Dependence of $f$ on degree of connectivity. At the lower and higher degrees, the plots are slightly separated though, around $\langle k \rangle=8$ they almost merge. The figures are obtained from numerical simulations on a system of size $25 \times 25$ under a 20-cycle scheme ($S_{(1)} = 0.79$). (b) The plot of (a) replotted on a semi-logarithmic scale, for clarity at high degrees.}
\end{figure}

\section{Degree distributions}
\label{sec:3}
Random wiring results in a Poissonian \cite{BAREV,ER1} distribution about the average degree $\langle k \rangle$, whereas the PA(I) scenario yields a Gaussian form around $\langle k \rangle$ \cite{BAREV,BBASI1}. For an $m$-cycle scheme and additional wiring probability $p$, the average $\langle k \rangle$ is given by, $$\langle k \rangle = 2mp+4,$$ as they all initially have degree 4. For example, if $m=20$, $$\langle k \rangle = 40 \times p+4.$$ The probability of a node to have degree $k$ is found by using, $$P(k) = \frac{n_k}{N},$$ where $n_k$ is the number of the nodes with degree $k$, and $P(k)$ is plotted as a function of $k$ in Fig. 1, giving the degree distribution curves. These distribution plots show a shift with their peaks about $\langle k \rangle$ at different $p$ values. We explain these as follows.

{\bf{Random:}} As all the nodes are randomly connected, most of them have degree $\langle k \rangle$. This leads to a relatively sharp peak about the average.

{\bf{PA(I):}} Here, due to the presence of hubs there will be more nodes at higher degrees; this should be compensated by the other nodes, as the total degree is conserved. Thus, there will be several nodes with degrees far away from $\langle k \rangle$ on both sides of the peak.

{\bf{PA(II):}} 
The process of PA(I) is heightened: here pairs of nodes with the highest connectivities are linked preferentially, leading to even higher fluctuations about the mean, and a corresponding flattening of the (area-preserving) curve.

%%%%%%%%%%%%%%%%%%%%%%%%%%%%%%%%%%%%%%%%%%%%%%%%%%%%%%%%%%%%%%%%%%%%%%%%%%%%%%%%%%%%%%%%%%%%%%%%%%%%5

\section{Survival probability}
\label{sec:4}

The variation of the survival probability $S(s)$ with $s$ at different $p$ in Stage II is shown in Fig. 2 for a 20-cycle scheme for scenarios corresponding to random wiring and PA(I). 
Though the same two-stage glassy \cite{JML} dynamics persist for both wiring strategies, the first, non-interacting stage of independent dynamics is fully unaffected by the choice of wiring strategy, resulting in the same value of $S_{(1)}$ for both cases; so it is not shown in the figure. In the second stage, an increase in the value of $p$ results, as expected, in a decrease in the number of survivors in both cases, where it is higher for PA(I) at a given $p$. 

\begin{figure}
\centering
\resizebox{0.6\columnwidth}{!}{
\includegraphics{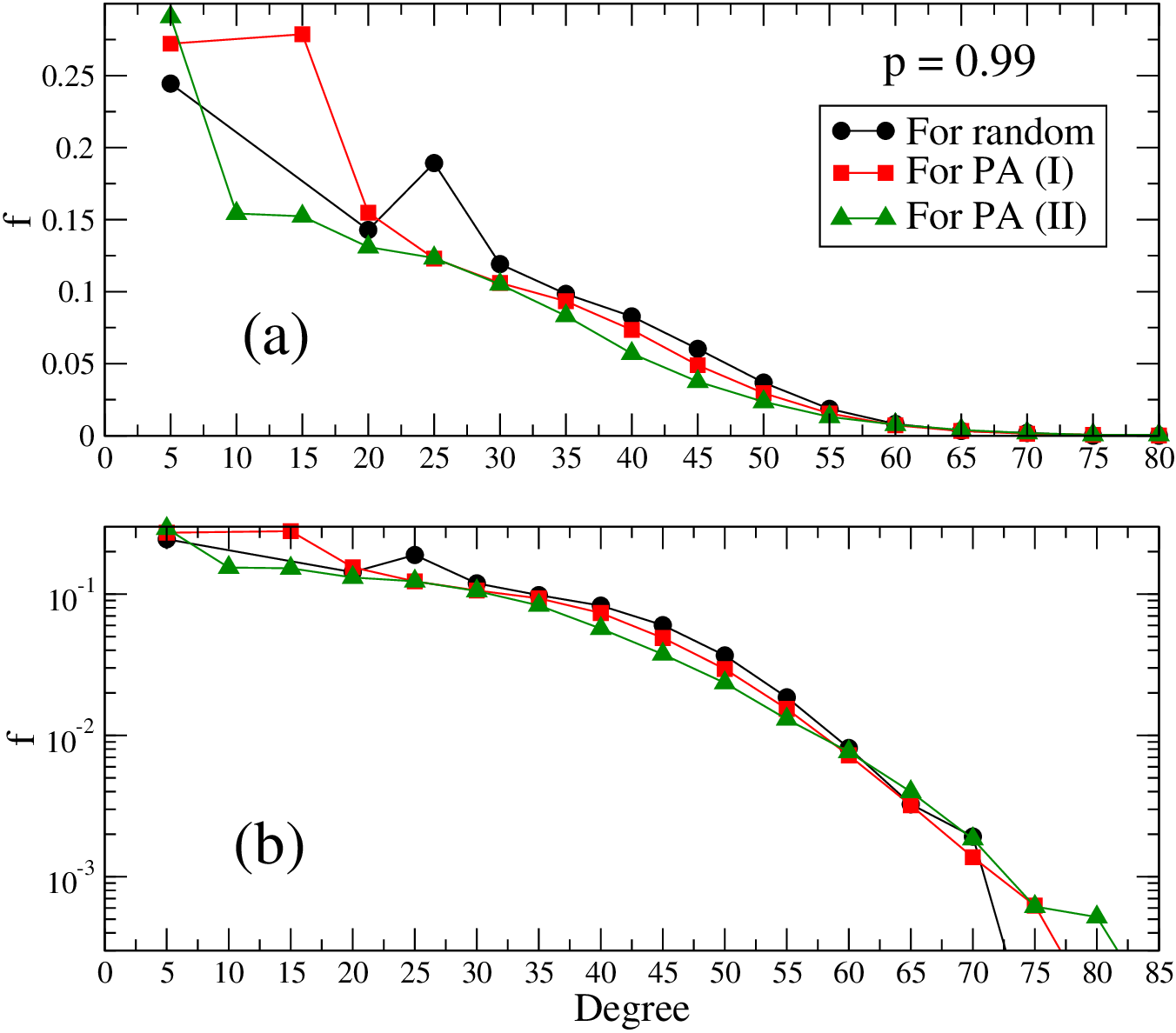} }
\caption{(Color online). (a) Dependence of $f$ on degree of connectivity. At low and high degrees, there are no distinct differences, but around $\langle k \rangle(\simeq 43)$ they are significantly separated. The figures are obtained from numerical simulations on a system of size $25 \times 25$ under a 20-cycle scheme ($S_{(1)} = 0.79$). (b) The plot of (a) on a semi-log scale. }
\end{figure}

%%%%%%%%%%%%%%%%%%%%%%%%%%%%%%%%%%%%%%%%%%%%%%%%%%%%%%%%%%%%%%%%%%%%%%%%%%%%%%%%%%%%%%%%%%%%%

\section{Dependence of survival on degree of nodes}
\label{sec:5}

Survivors on a regular lattice become vulnerable as they are connected to more and more neighbours via wiring. We investigate that for the three wiring scenarios in this section. If the number of the nodes in a given interval of degree is $\Gamma$ and the number of the eventual survivors in that interval is $\Sigma$, the survival fraction $f$ is found by using, $$f=\frac{\Sigma}{\Gamma}.$$ In Figs. 3 and 4, we bunch degrees into bins of five, and plot $f$ with respect to degree. For all the wiring methods, we note that $f$ falls off sharply at higher degrees.

For low $p$, we notice that the different wiring scenarios give rise to slight differences only for low and high degrees of connectivity. Actually at low p, the few additional edges are distributed among all the nodes uniformly for random case and so the survival of the clusters for all degrees are equally affected. But, preferential attachment causes formation of hubs and so the effect on survival of the clusters on their degree is noticed at high degree (for hubs) and at low degree (for poorly connected nodes), leading to differences in $f$ values for three scenarios at the ends . 

These remarks for the low- and high-degree ends of the plot apply also to the large $p$ case (Fig 4). Here, however, the curves for the different wiring scenarios are also well separated around their middle. Contrary to the low-$p$ case, the formation of hubs is now complete, so that the topologies in the three cases are rather different. The random wiring scenario has the largest fraction of survivors, consistent with its having the maximum number of nodes around $\langle k \rangle$ (Fig. 1, $p=1.0$), the reduced value of which decreases the fraction of survivors in the PA(I) and PA(II) cases. 

\begin{figure} 
\centering
\resizebox{0.55\columnwidth}{!}{
\includegraphics{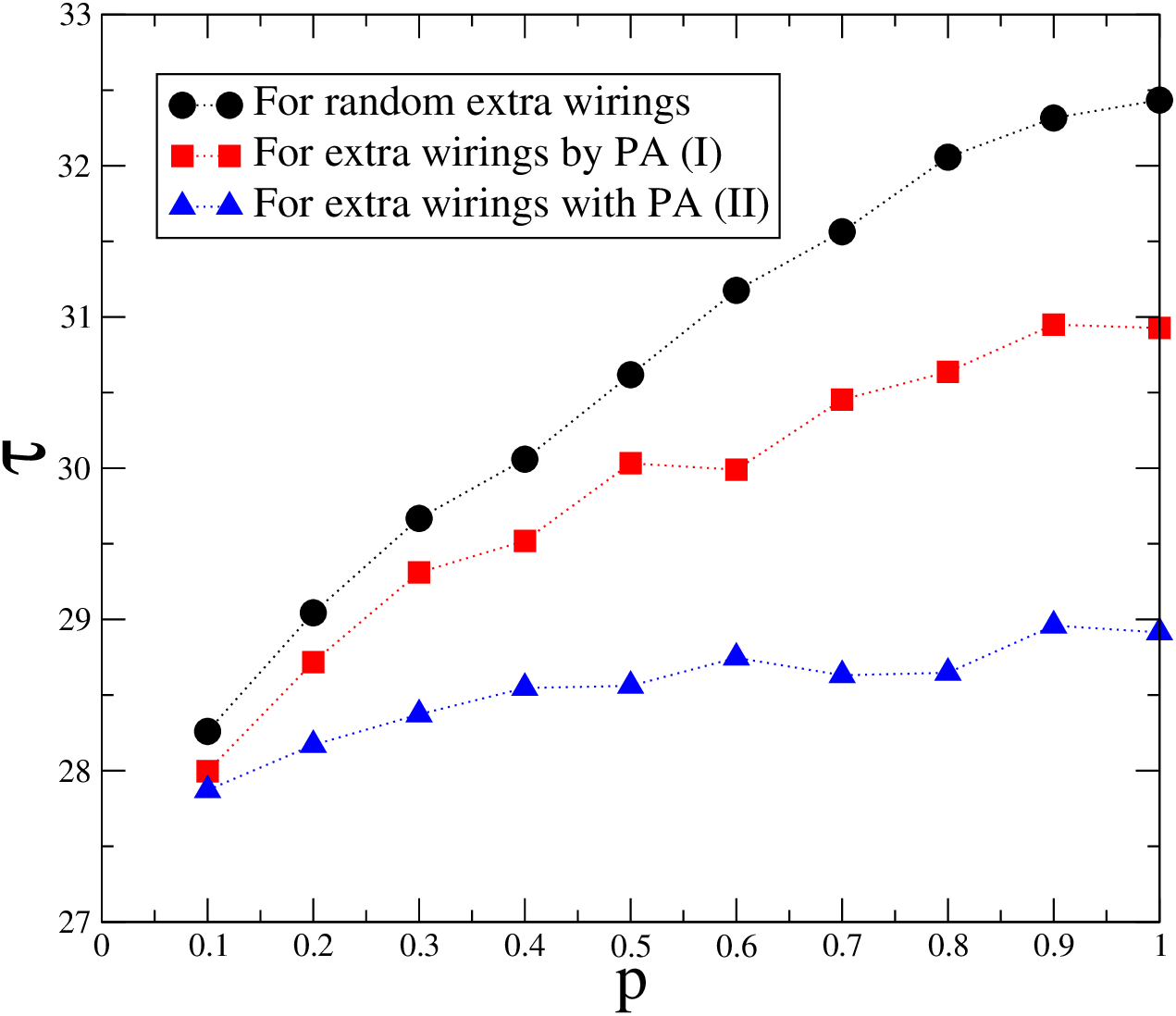} }
\caption{(Color online). 
Plot of the variation of $\tau$ with $p$ in the three wiring scenarios referred to in the text, obtained from numerical simulations on a system of size $25 \times 25$ under a 20-cycle scheme ($S_{(1)} = 0.79$). The value of $\tau$ is maximal for the random, minimal for PA(II), and intermediate for PA(I). 
}
\end{figure} 
%%%%%%%%%%%%%%%%%%%%%%%%%%%%%%%%%%%%%%%%%%%%%%%%%%%%%%%%%%%%%%%%%%%%%%%%%%%%%%%%%%%%%%%%%

\section{Time required for isolation of eventual survivors}
\label{sec:6}

As a result of the interactions shown in Eq. (3), eventual survivors will be isolated i.e., all their neighbours have to die out first.
The time required for four nearest neighbours on a square lattice to die is clearly less than that for the far greater number of neighbours in a wired network. Thus, the reduced time $\tau$ required for the isolation of the ultimate survivors should be longer for larger values of $p$. This is clearly seen in Fig. 5.

At a given value of $p$, the same number of total connections is added to the nodes in all three scenarios; also higher degree clusters (hubs) eventually die out, irrespective of their initial masses and the process of wiring. Keeping these facts in mind, the difference between the individual plots in Fig. 5 for different strategies of wiring can be explained as follows:

{\bf{Random:}} On average each of the nodes has $\langle k \rangle$ neighbours (which may be survivors or non-survivors with equal probability). Thus, eventual survivors have to wait for $\langle k \rangle$ neighbours to die out in order to get isolated.

{\bf{Preferential attachment:}} For both PA(I) and PA(II), hubs form, and eventually die out. This has two effects: first, non-hub sites can stay effectively isolated and survive forever. Second, the rapid death of hubs causes holes to appear and assist in the isolation of their neighbouring nodes. Both of these lead to a rapid isolation of eventual survivors, which is faster for PA(II) than PA(I), leading to the behaviour presented in Figure 5.  

%%%%%%%%%%%%%%%%%%%%%%%%%%%%%%%%%%%%%%%%%%%%%%%%%%%%%%%%%%%%%%%%%%%%%%%%%%%%%%%%%%%%%%%%%%%%%%%%%%%%%5

\section{Eventual survivors for high p}
\label{sec:7}

It is interesting to examine the behaviour of the number of eventual survivors with respect to wiring strategies for fully developed networks (high $p$).
The number of eventual survivors, given by $N_{ev}$, is reduced at higher $p$, irrespective of the method of wiring (Fig. 6). The plots can be fitted to $N_{ev} = a+b$ $ln(p),$ where $a$ and $b$ are arbitrary constants and depend on the wiring scenario (these values are provided in Fig. 6 separately for the three processes). For a particular value of $p$, $N_{ev}$ is maximal for PA(II) and minimal for the random case. This can be explained as follows:

\begin{figure} 
\centering
\resizebox{0.55\columnwidth}{!}{
\includegraphics{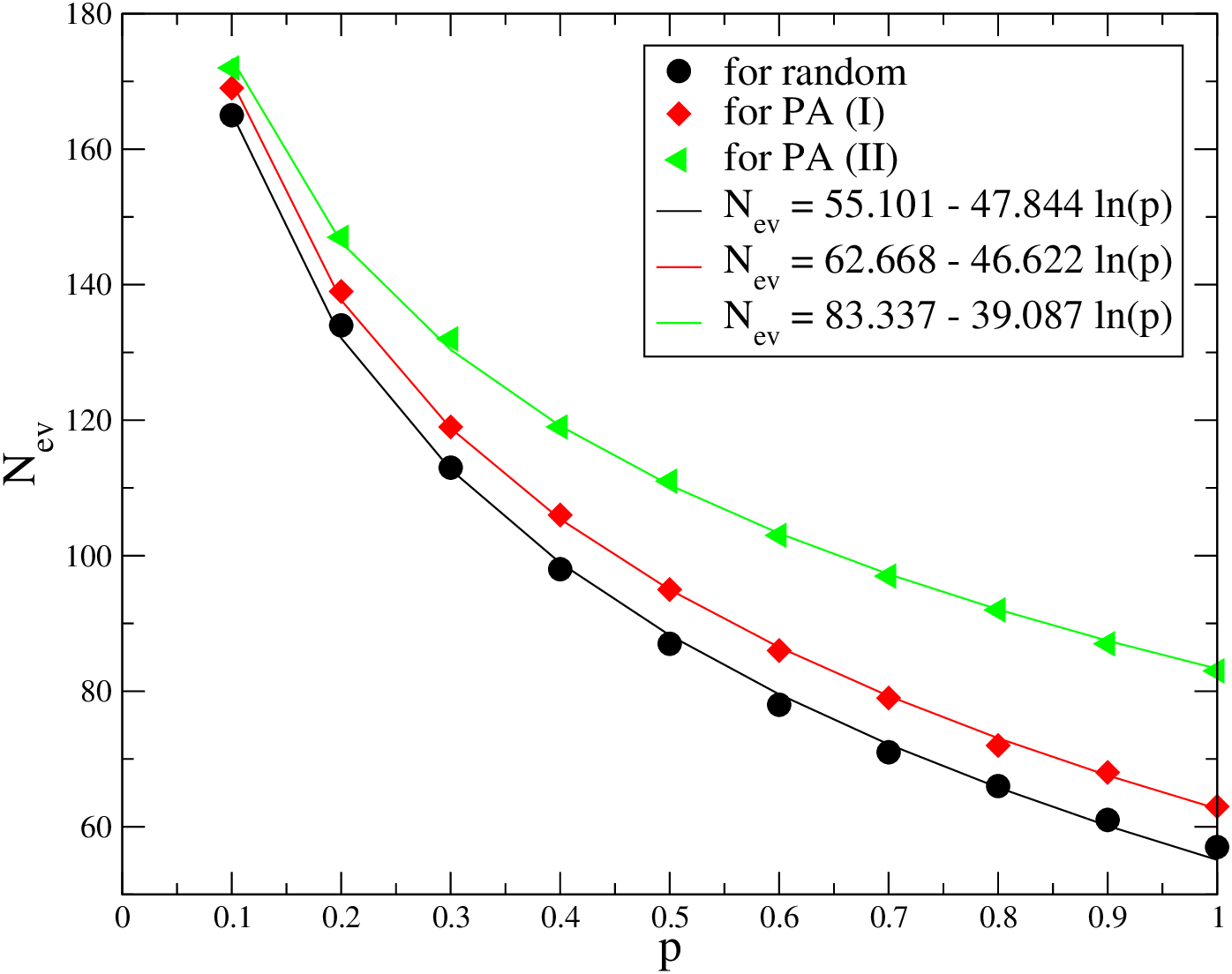} }
\caption{(Color online). Variation in the total number of the ultimate survivors with $p$ for three wiring methods. The solid lines are governed by the equation $N_{ev} = a+b$ $ln(p)$, with the different $a$ and $b$ values for different wiring methods (as indicated). The plots are obtained from numerical simulations on a system of size $25 \times 25$ for a 20-cycle scheme ($S_{(1)} = 0.79$).}
\end{figure}

{\bf{Random:}} The clusters which were eventual survivors in the regular lattice are equally likely to be connected to all other clusters, irrespective of whether these are asymptotic survivors or not. This results in $N_{ev}$ decreasing sharply with increase in $p$.

{\bf{PA(I):}} 
Hubs, which eventually die out, monopolize the connectivity, leaving many non-hub sites effectively isolated, as potential survivors. This results in a larger value of $N_{ev}$ for a given $p$ than the random case.

{\bf{PA(II):}} In this case, hubs are preferentially interconnected and there is even less competition among the other nodes than there is for PA(I), resulting in an even larger number of potential non-hub survivors.

%%%%%%%%%%%%%%%%%%%%%%%%%%%%%%%%%%%%%%%%%%%%%%%%%%%%%%%%%%%%%%%%%%%%%%%%%%%%%%%%%%%%%%%%%%%%%%%%%%

\section{The fate of clusters with initial masses $x<x_*$}
\label{sec:8}

As noted previously \cite{JML}, two-stage dynamics are given by Eq. (3).  In the fast dynamics of Stage I, clusters with initial masses less than some threshold $x_*$ die out without interactions among themselves: hence the precise strategy of additional wiring has no effect.  We here examine the effect of initial mass distributions on the isolation time $\tau$, and the number of eventual survivors $N_{ev}$. In the first case, masses are distributed in a $25\times25$ system without restriction, while in the second case, only above-threshold masses with $x>x*$ are included. Our results for $\tau$ and $N_{ev}$ for different $p$ values are plotted in Figs. 7 and 8 respectively.

All above-threshold masses are would-be survivors, which results in a competition of survival among more clusters than in the unrestricted case. This results in more competition, and hence a larger $\tau$ for a given $p$, irrespective of the wiring strategy (Fig. 7). For similar reasons, the situation where all masses are above threshold will generate more potential, and hence more eventual survivors for all the wiring strategies (Fig. 8).

\begin{figure} 
\centering
\resizebox{0.55\columnwidth}{!}{
\includegraphics{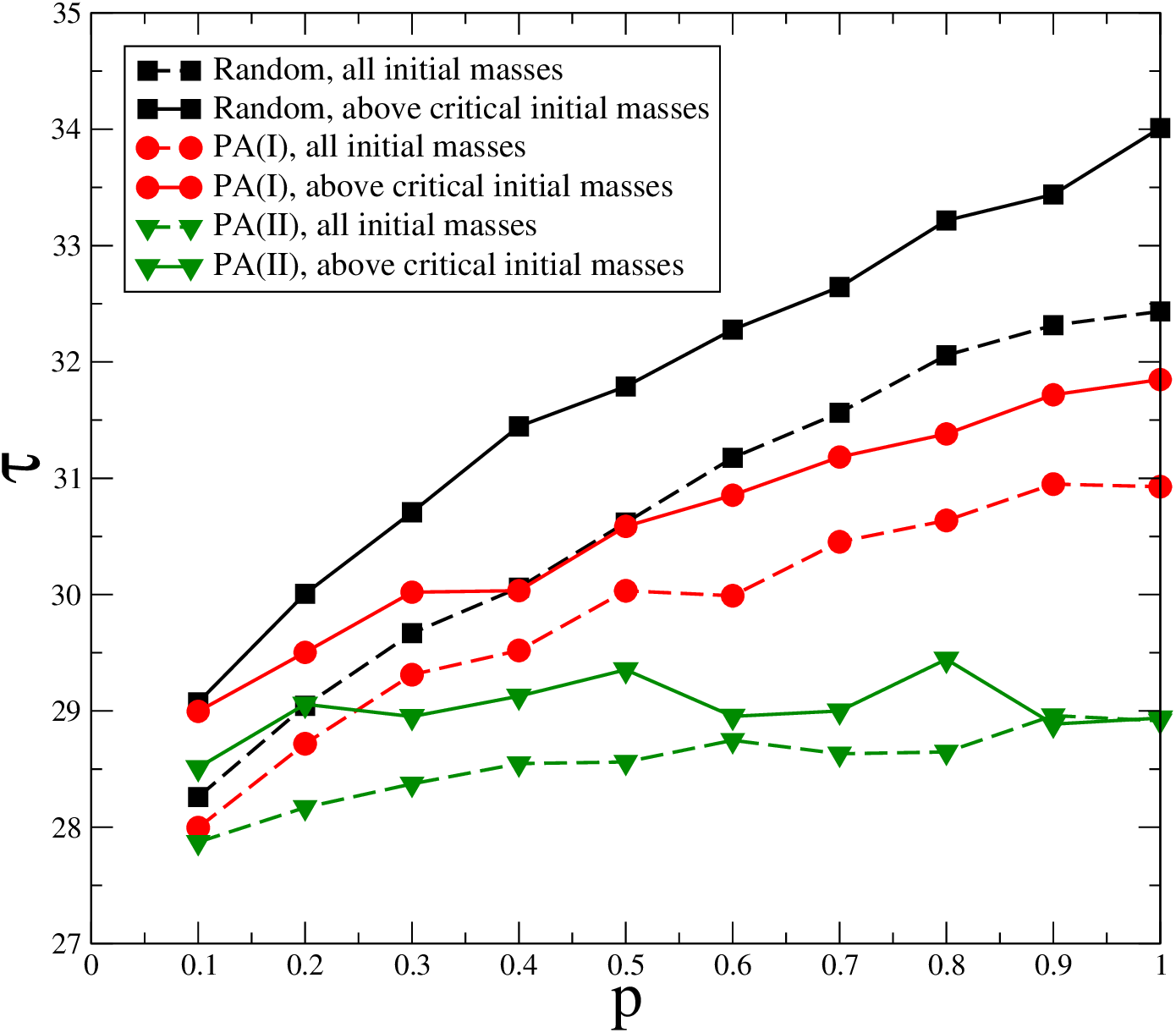} }
\caption{(Color online). The comparison of the value of $\tau$ in a $25 \times 25$ system under the 20-cycle scheme in three strategies of extra wiring. The plots with the point-dashed line are for the lattice having initial all masses corresponding to $S_{(1)}=0.79$ and the point-solid line plots are for that system, in which the clusters are with their initial masses corresponding to $S_{(1)}=0.79$ but, excluding the masses less than $x_* ( = \sqrt{2})$ (i.e., with all would-be survivors). So, the latter case offers more competition among the clusters raising the $\tau$ value for a particular $p$ under any process of wiring.}
\end{figure}

%%%%%%%%%%%%%%%%%%%%%%%%%%%%%%%%%%%%%%%%%%%%%%%%%%%%%%%%%%%%%%%%%%%%%%%%%%%%%%%%%%%%%%%%%%%%%%%%%%%%%%%%%%%%%%%%5

\section{Cumulative mass difference distributions for different wiring scenarios}
\label{sec:9}

\subsection{Cumulative mass difference distributions for all clusters}
\label{sec:10}
For a regular square lattice, every node has four nearest neighbours with which it interacts. But, with extra wiring on the lattice, this number of neighbours varies from node to node and the cumulative mass difference is found by using:

\begin{equation}
\sum \delta x_i = k_i x_i - \sum_{j=1}^{k_i}x_j,
\end{equation} 
for the $i$-th node, where $k_i$ is its degree and $x_i$ is its initial mass.

The plots of the cumulative mass difference distribution for all clusters are shown in Fig. 9 for different wiring strategies. With increasing $p$, plots are increasingly flat at the centre. 
For a regular lattice, masses are typically homogeneously distributed, so that the cumulative mass difference is sharply peaked around zero. As the number of neighbours increases with increasing $p$, the distribution gets progressively wider. For the random wiring case, this leads to a progressive flattening, since cumulative mass differences are distributed relatively homogeneously around zero. This is emphatically not the case for PA(I) and PA(II) which represent increasing inhomogeneity of cluster neighbourhoods in terms of masses, and lead additionally to wide tails in the distribution.

%%%%%%%%%%-------------------------------------------------------------

\subsection{Cumulative mass difference distributions for eventual survivors}
\label{sec:11}

Using Eq. (4), $\sum \delta x_i$ is found for only the eventual survivors under different extra wiring processes, as shown in Fig. 10. This plot gives rise to one of the most important observations of this paper. 
First we note that preferential attachment makes little difference to the low $p$ limit, which reinforces the findings of ref. \cite{NNT} in its clear indication of survivors against the odds. The portion of the plots to the left of zero represents  masses which were potential `losers' (with masses less than the cumulative sum of their neighbours), but which became eventual survivors. As $p$ increases, the effect of preferential attachment becomes clearly visible by 
resulting in the highest peak for PA(II) due to a large fluctuation in the average neighbour masses, the reduction of which gives the opposite consequence for random scenario, with PA(I) as their intermediate.

\begin{figure}
\centering
\resizebox{0.55\columnwidth}{!}{
\includegraphics{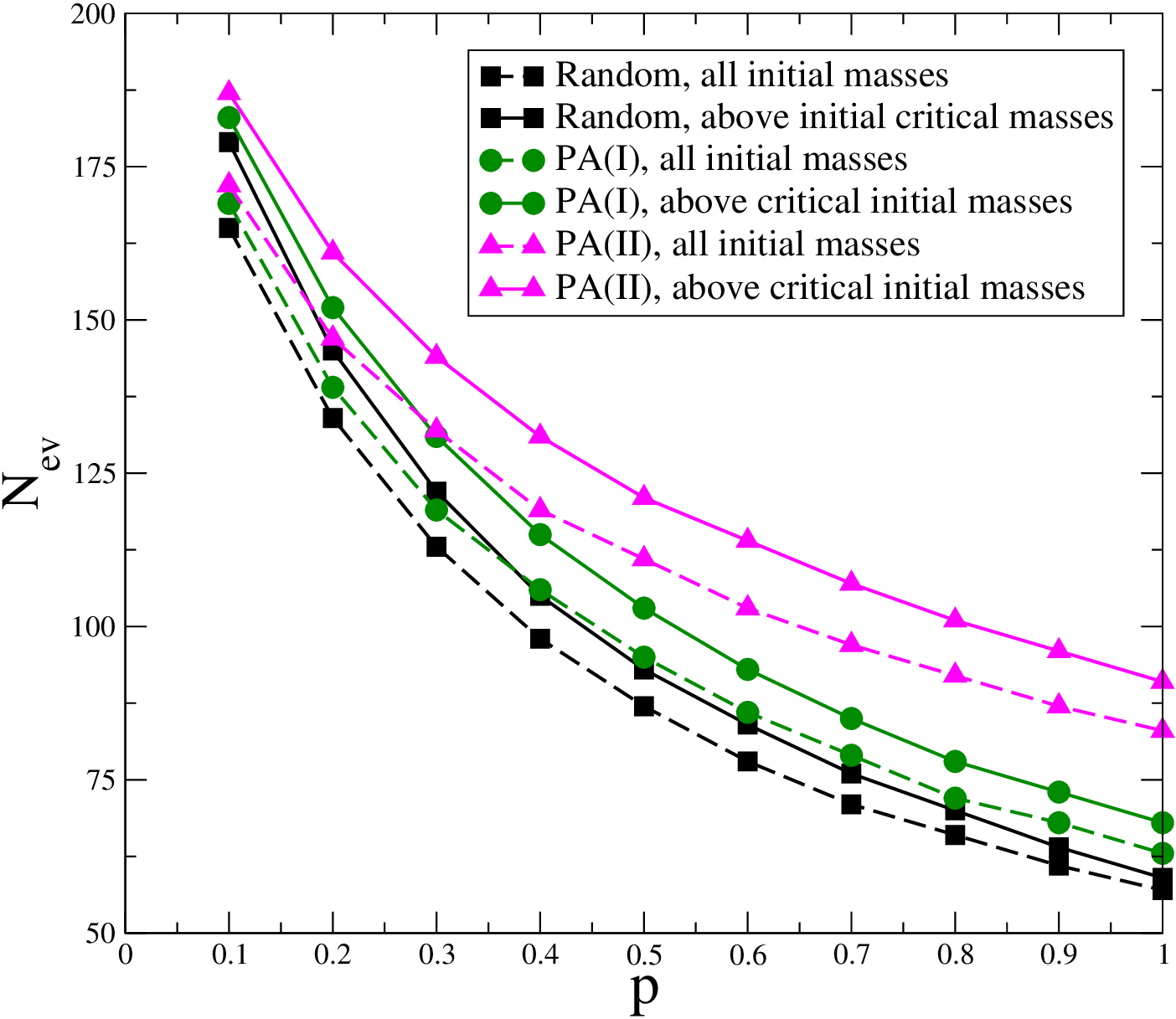} }
\caption{(Color online). The comparison of the number of the eventual survivors in a $25 \times 25$ system with a 20-cycle scheme for the three wiring strategies. The dot-dashed lines and dot-solid lines correspond to $S_{(1)}=0.79$, with the difference that in the latter case all masses less than threshold are excluded, thus increasing the number of ultimate survivors.
}
\end{figure}

Also, while random wiring increases the relative number of potential winners among the eventual survivors, the preferential attachment scenarios do exactly the opposite (Fig. 10). Fig. 11 compares the plots of Fig. 10 with those of Fig. 9. From the figure it is clear that, with the rise in $p$ the separation at the right hand end of the dashed-line plot and the solid-line plot for a given scenario gradually rises, implying a more uncertainty for even the highly massive cluster to become the eventual survivors at a higher $\langle k \rangle$.

Finally, in this section we direct our attention to winners against the odds  \cite{NNT}, that is those clusters with masses on the negative side of zero in Fig. 10. The ratio of the area under the latter to the total area of the curve is denoted by $\alpha_{ev}$ and plotted as a function of $p$ in Fig. 12. Preferential attachment significantly enhances $\alpha_{ev}$, and thus the chances of winning against the odds. The simple reason behind this is that preferential attachment links erstwhile massive high-degree clusters with each other, leading to their mutual annihilation, while the `little guys' of the non-hub sites are left alone with fewer neighbours and consequently a greater chance of survival. 

\section{Discussion}
\label{sec:12}

We here summarize the main results. Random wiring leads to the formation of a relatively homogeneous network, while the preferential attachment processes produce hubs. These hubs are the strongest (i.e., with the highest degree values) for PA(II) case so that many of the nodes of the network remain poorly connected (Fig. 1).

The homogeneity leads to a relatively uniform spread of competition among clusters and gives rise to a minimum for $N_{ev}$ and a maximum for $\tau$ in the case of random wiring. 
The strong inhomogeneities associated with hubs lead to the opposite scenario for PA(II) and an intermediate one for PA(I) (Fig. 5,6).

Extra wiring leads to the most massive clusters becoming vulnerable, irrespective of the specific strategy used (Fig 11). The effect of preferential attachment is extremely significant here, as it promotes erstwhile losers into ultimate survivors (while eliminating earlier massive potential survivors) in a pronounced way, with respect to the random wiring strategy. This is easily seen from Fig 12, where an increase in $p$ leads to an increase in $\alpha_{ev}$ for PA(II) and a decrease in this for the random case.

In the context of economics, this has an interesting implication. Contrary to public perceptions, the entry of large conglomerates to the commercial sector and their consequent cutthroat competition could, paradoxically, lead to the survival of corner shops, which would be isolated as `non-hubs' in the marketplace.

Of course this claim is contingent on preferential attachment: that is, the conglomerates should trade with each other, rather than with corner shops. A large retailer which purchases wholesale from a major corporate brand is thus more vulnerable to extinction than one that picks up its supplies from a range of outlets independent of their size. In cases such as the former, the larger players are likely to die out, while the corner shops survive.

~\newpage
\begin{figure} 
\centering
\resizebox{0.634\columnwidth}{!}{
\includegraphics{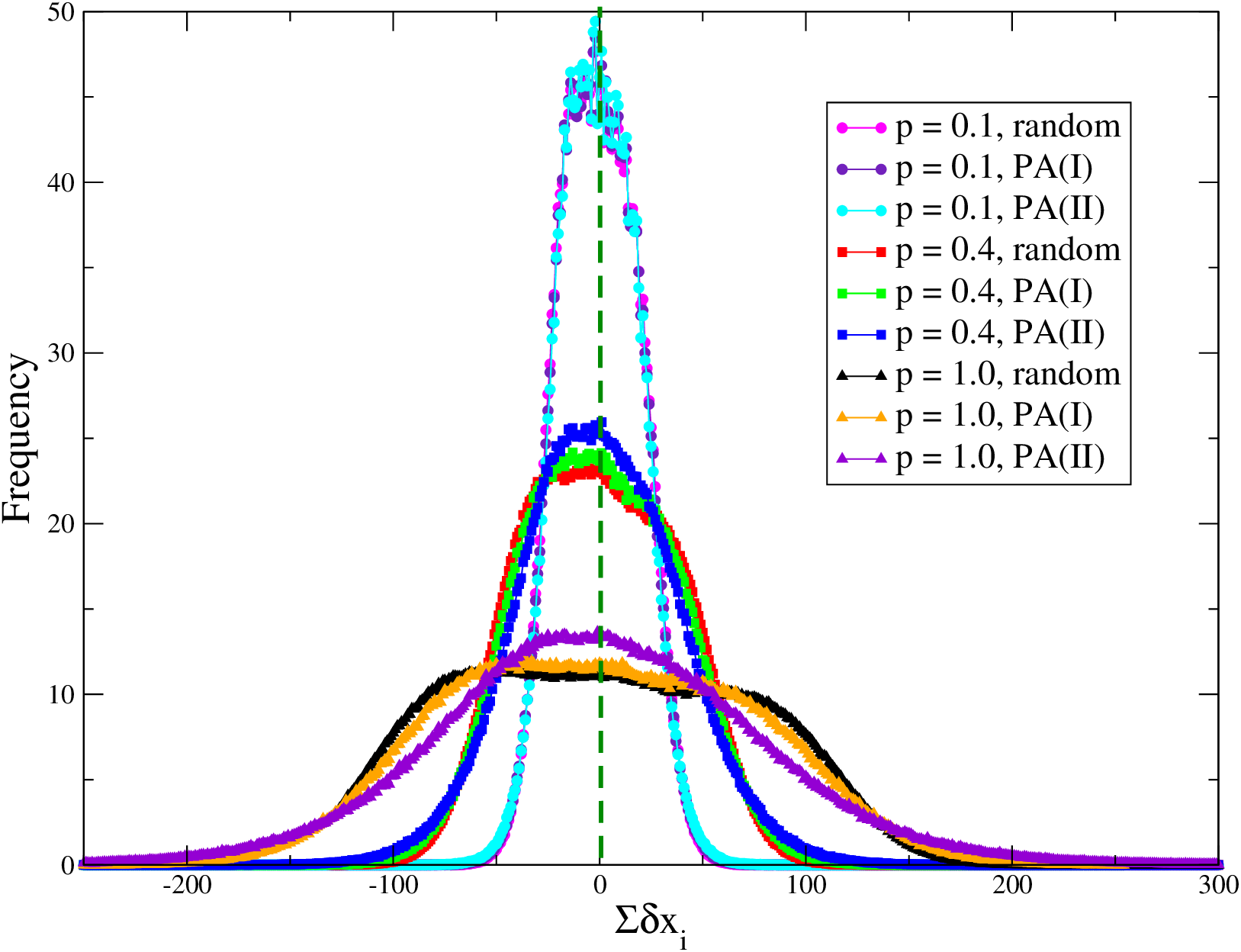} }
\caption{(Color online). The cumulative mass difference distributions for all clusters due to three wiring processes in a $50 \times 50$ lattice under the 10-cycle scheme ($S_{(1)} = 0.85$). At a larger $p$, the curve shows more flattening at the central region due to a larger centralization of the average neighbour mass for each cluster. For a given $p$, the centralization is maximum for the random case and minimum for PA(II), resulting in the highest central value of the curve for PA(II), the opposite for the random case, and for PA(I) the corresponding curve is intermediate to them. }
\end{figure}

\begin{figure}
\centering
\resizebox{0.634\columnwidth}{!}{
\includegraphics{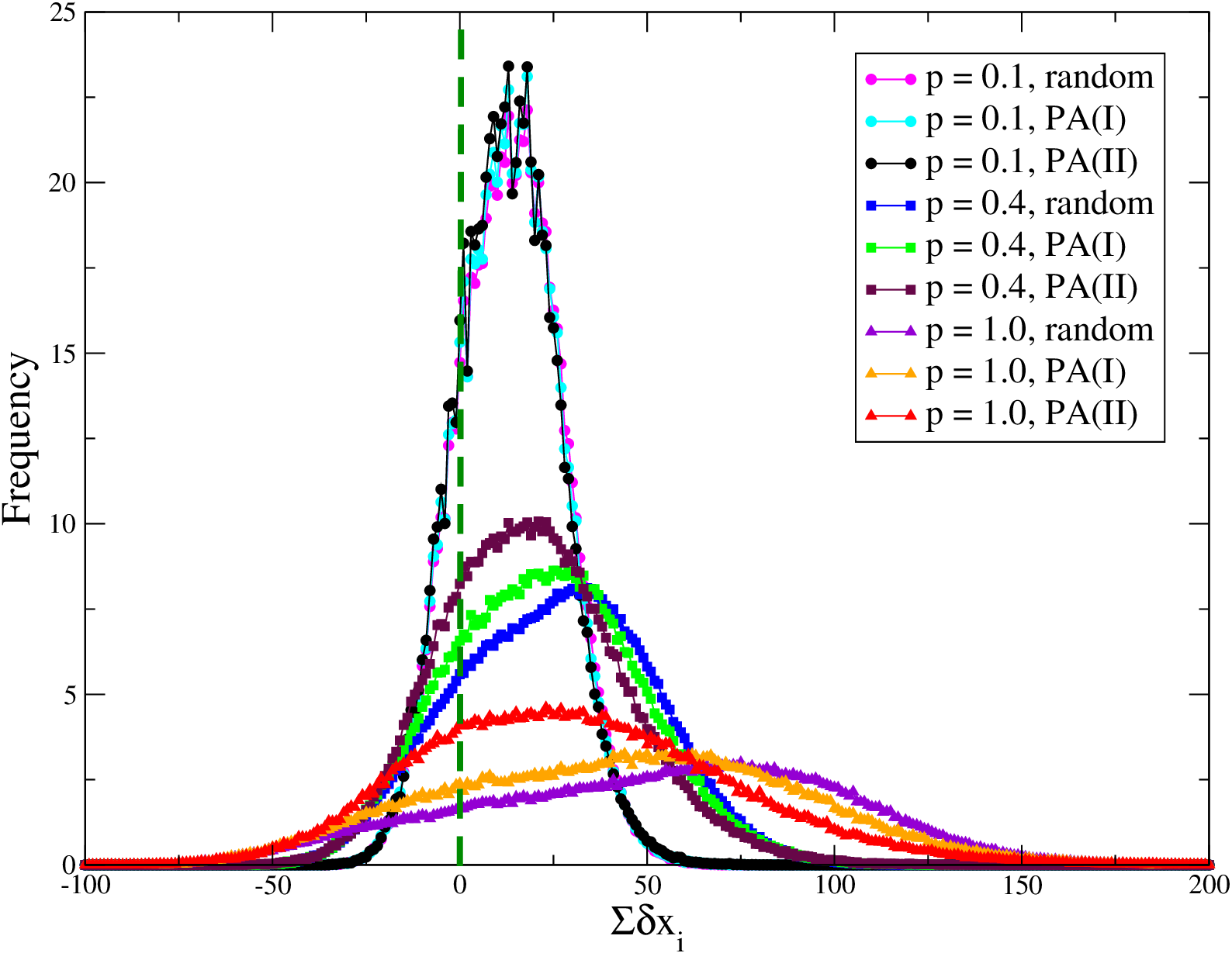} }
\caption{(Color online). The cumulative mass difference distributions for the eventual survivors in three wiring processes. The area under the negative portion of the curve at any $p$ is less than that under the positive portion implying a larger surviving possibility for the massive clusters. The curve for a given $p$ has the highest peak for PA(II) and the lowest peak value for random case. The plots are obtained numerically from the same $50 \times 50$ lattice of Fig. 9 under the 10-cycle scheme due to all scenarios of additional wiring ($S_{(1)} = 0.85$). }
\end{figure}

\begin{figure} 
\centering
\resizebox{0.6\columnwidth}{!}{
\includegraphics{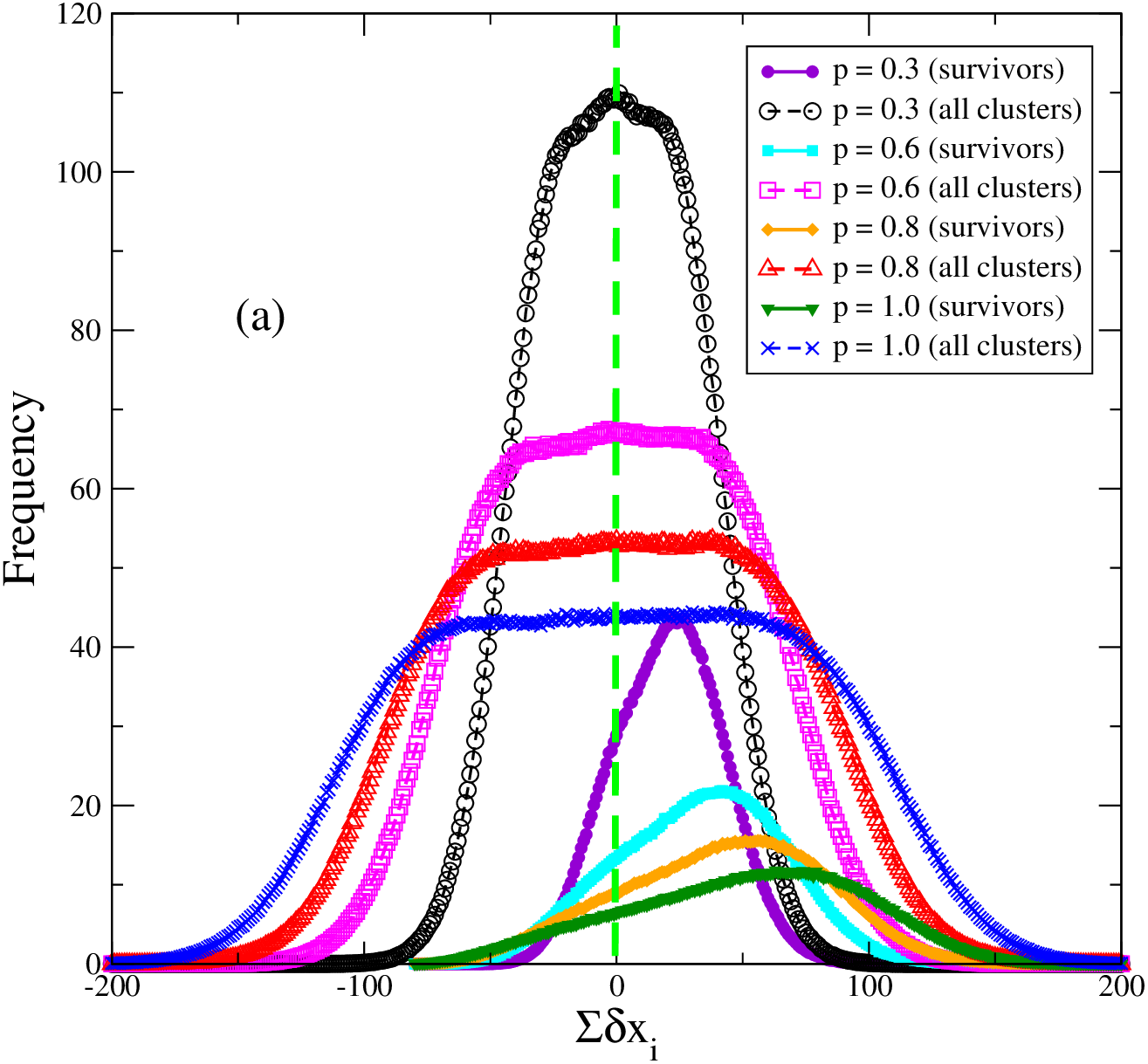}}

\resizebox{0.6\columnwidth}{!}{
\includegraphics{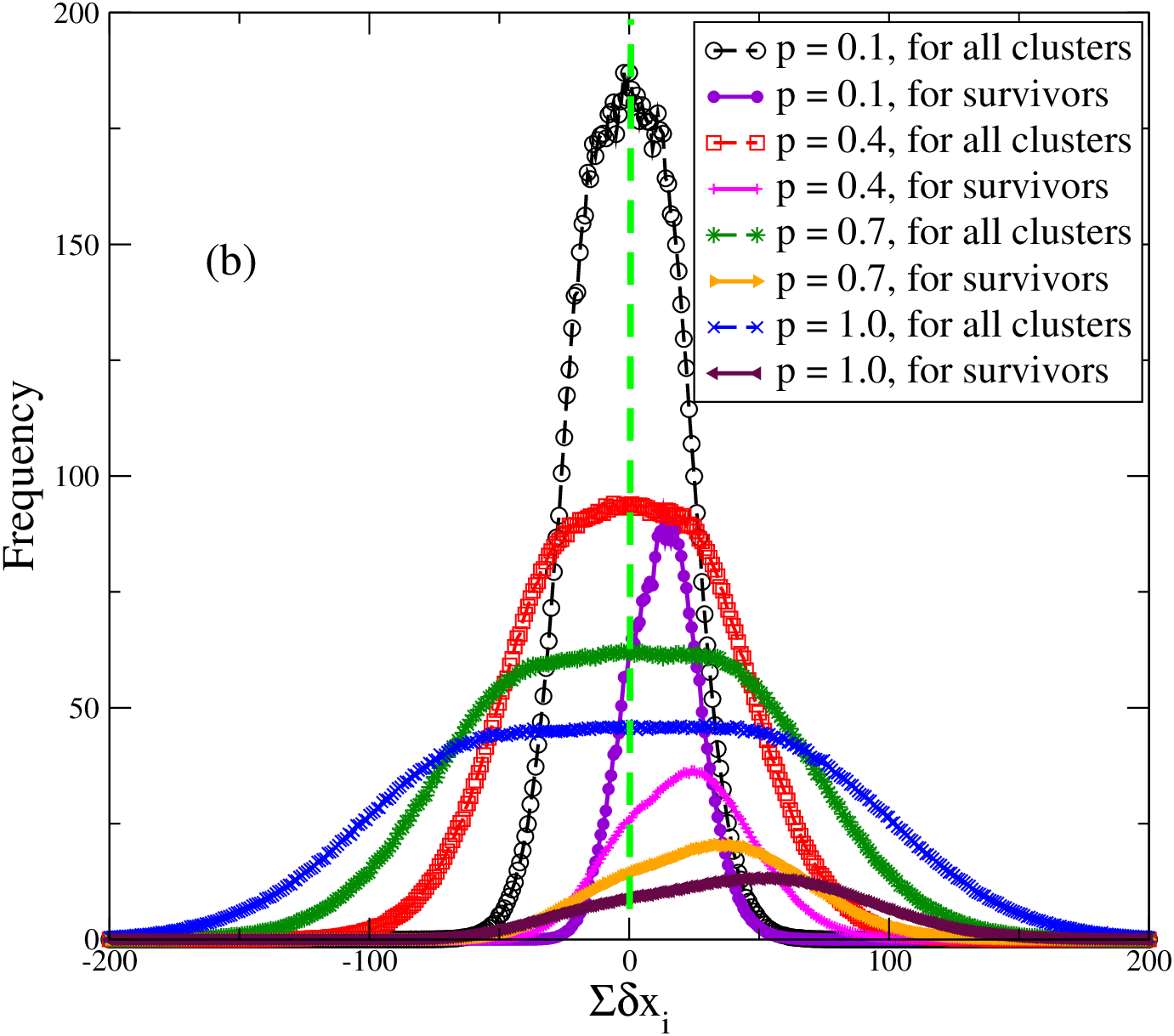}}

\resizebox{0.6\columnwidth}{!}{
\includegraphics{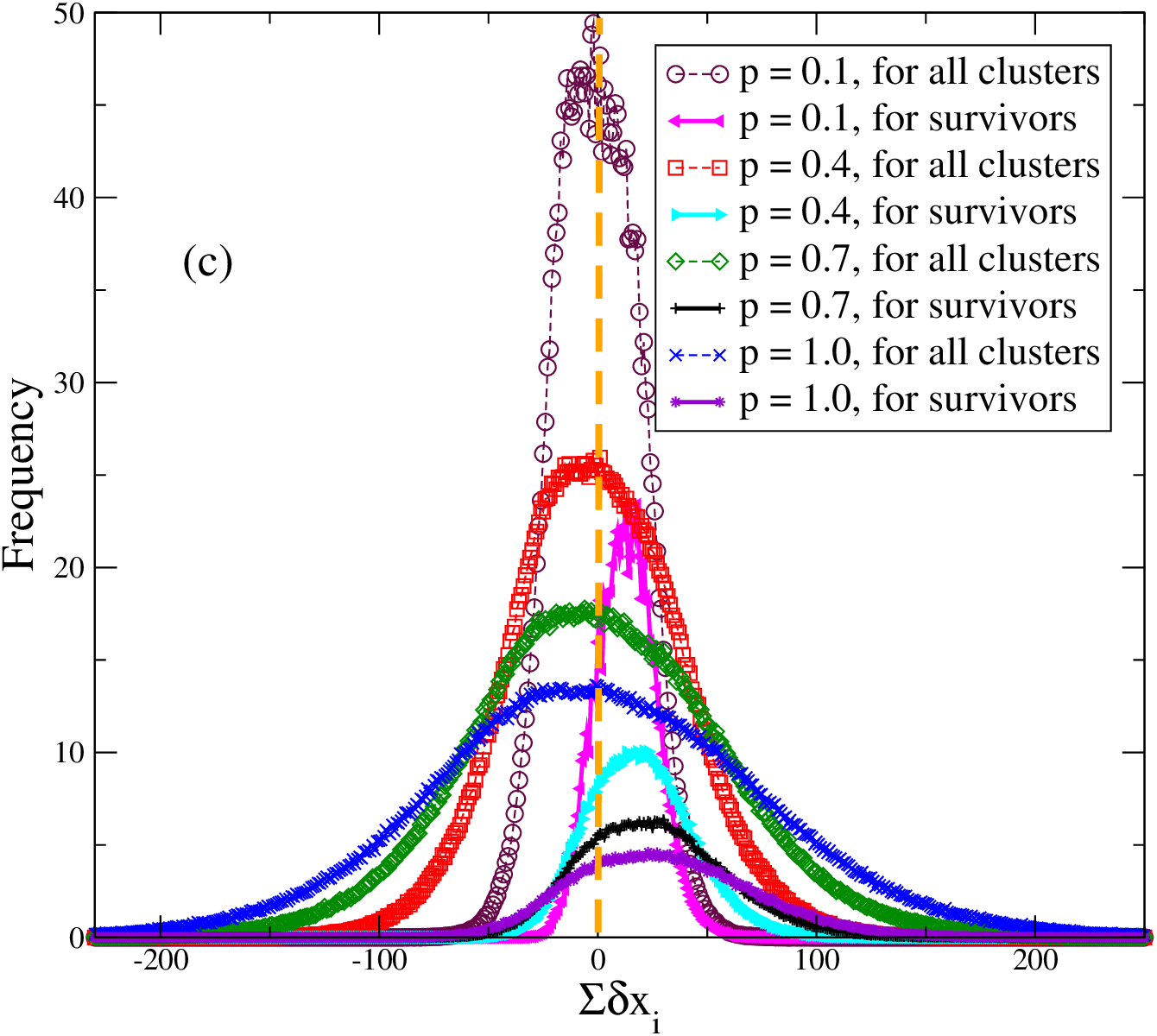}}
\caption{(Color online). The comparison of the cumulative mass difference distributions for all clusters and for the eventual survivors under the - (a) random, (b) PA(I) and (c) PA(II) wiring processes in a $100 \times 100$ system for the first two cases and in a $50 \times 50$ system for the third one at different $p$ values, as indicated ($S_{(1)} = 0.85$).   }
\end{figure}

%%%%%%%%%%%----------------------------------------------------------------------

\begin{figure}
\centering
\resizebox{0.55\columnwidth}{!}{
\includegraphics{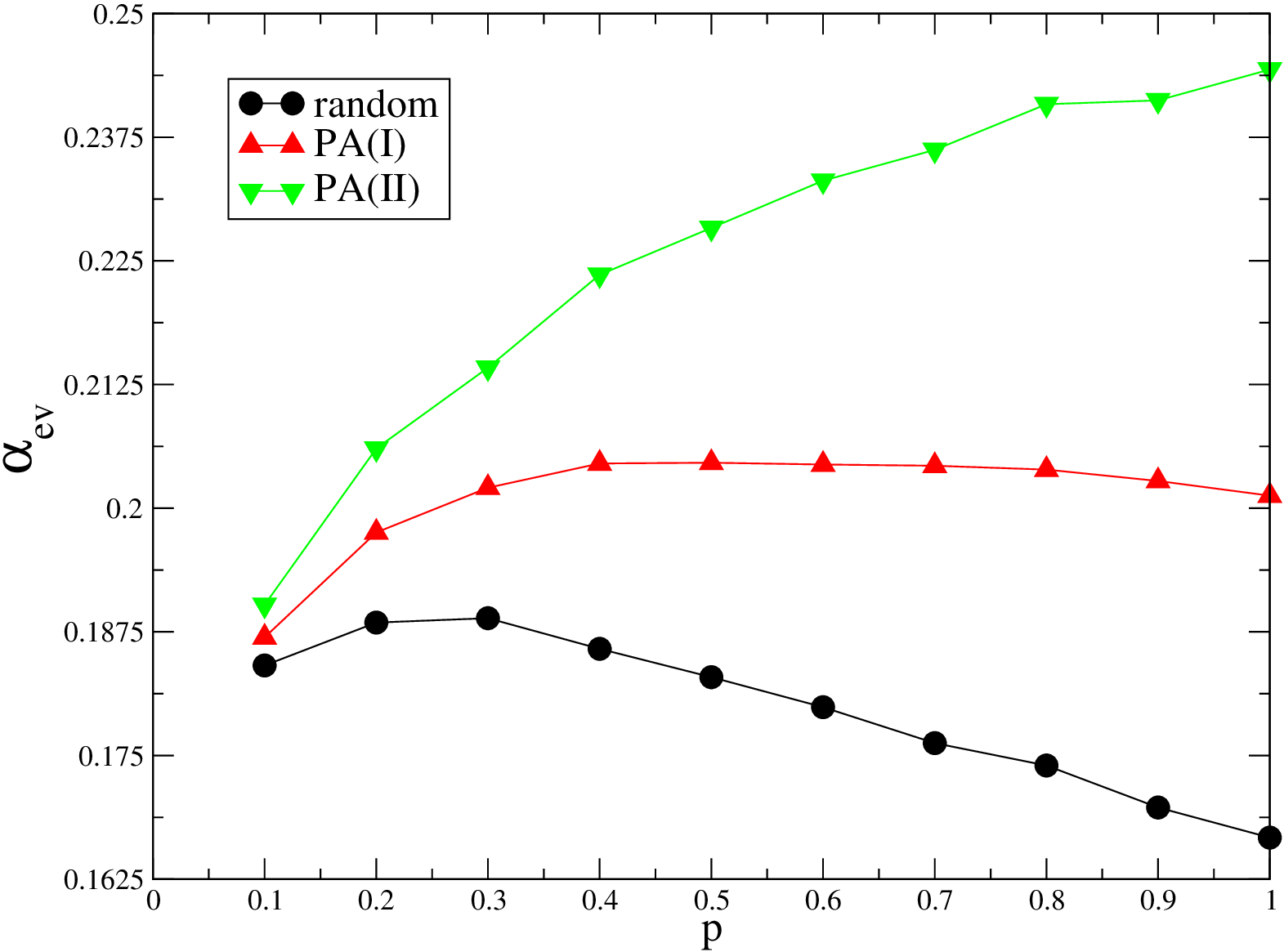} }
\caption{(Color online). Variation of the fraction of survivors against the odds, $\alpha_{ev}$, for different wiring strategies in a system of size $50 \times 50$ ($S_{(1)} = 0.85$). For a given $p$, preferential attachment greatly increases the number of survivors against the odds. 
}
\end{figure}

%~\newpage
~

\end{document}